\documentclass[lettersize,journal,review]{IEEEtran}

\usepackage{array}
    \newcolumntype{P}[1]{>{\raggedright\arraybackslash}p{#1}}
    \newcolumntype{M}[1]{>{\raggedright\arraybackslash}m{#1}}
    
\usepackage[caption=false,font=normalsize,labelfont=sf,textfont=sf]{subfig}
\usepackage{textcomp}
\usepackage{stfloats}
\usepackage{url}
\usepackage{verbatim}
\usepackage{graphicx}
\usepackage{cite}
\usepackage{enumitem}
\usepackage{multirow}
\usepackage{xtab}
    \xentrystretch{-0.1}

\usepackage{float}

\usepackage{xcolor}
\usepackage{soul}
\usepackage{amsmath}

\usepackage[numbers]{natbib}

\def\BibTeX{{\rm B\kern-.05em{\sc i\kern-.025em b}\kern-.08em
    T\kern-.1667em\lower.7ex\hbox{E}\kern-.125emX}}
\usepackage{balance}

\usepackage{graphicx}
\usepackage{textcomp} 
\usepackage{listings}
\usepackage{xcolor}

\lstdefinestyle{mystyle}{
    backgroundcolor=\color{white},   
    commentstyle=\color{gray},
    keywordstyle=\color{blue},
    numberstyle=\tiny\color{gray},
    stringstyle=\color{red},
    basicstyle=\ttfamily\footnotesize,
    breaklines=true,                     
    captionpos=b,                        
    numbers=left,                        
    numbersep=5pt,                       
    showspaces=false,                    
    showstringspaces=false,
    showtabs=false,                      
    tabsize=3,
    frame=single,
    morekeywords={let},
    rulecolor=\color{black},
    literate={~}{{\textvisiblespace}}1,  
    xrightmargin=0cm
}

\definecolor{interfaceblue}{rgb}{0.15, 0.54, 0.68}
\definecolor{caribbeangreen}{rgb}{0.0, 0.8, 0.6}
\definecolor{codegreen}{rgb}{0,0.6,0}
\definecolor{commentgray}{rgb}{0.5,0.5,0.5}
\definecolor{keywordblue}{rgb}{0.0,0.0,1.0}
\definecolor{structgreen}{rgb}{0.22,0.75,0.69}
\definecolor{lightblue}{HTML}{DAE8FC}
\definecolor{lightorange}{HTML}{FFE6CC}
\definecolor{lightred}{HTML}{F8CECC}
\definecolor{lightgreen}{HTML}{D5E8D4}
\lstdefinelanguage{CSharp}{ 
  language=[Sharp]C,
  basicstyle=\ttfamily\footnotesize,
  keywordstyle=\color{keywordblue},
  commentstyle=\color{commentgray},
  showstringspaces=false,
  breakatwhitespace=false,
  breaklines=true,
  tabsize=2,
  frame=single,
  morekeywords={where, new, var},
  moredelim=[is][\color{interfaceblue}]{\^}{\^},
  moredelim=[is][\color{keywordblue}]{\#}{\#},
  moredelim=[is][\color{structgreen}]{\%}{\%},
  xleftmargin=1.5em,
}

\newtheorem{definition}{Definition}

\begin{document}

\title{Bidirectionalization For The Common People}

\author{
Juraj Dončević \IEEEmembership{Member, IEEE}, 
Mario Brčić \IEEEmembership{Member, IEEE}, 
Danijel Mlinarić \IEEEmembership{Member, IEEE}

\thanks{J. Dončević, M. Brčić and D. Mlinarić are with the University of Zagreb Faculty of Electrical Engineering and Computing, Zagreb, Croatia.}
}

\markboth{February 2025}%
{}

\maketitle

\begin{center}
    \textbf{\textcolor{red}{This work has been submitted for possible publication. Copyright may be transferred without notice, after which this version may no longer be accessible.}}
\end{center}

\begin{abstract}
This paper presents an innovative approach to applying bidirectional transformations (BX) in practice. To introduce BX to a wider audience of technologists, engineers, and researchers, we have chosen to use C\# to develop \textit{Bifrons} - a library of BX lenses that replaces domain-specific programming languages (DSL) in practical use. The proposed approach simplifies the implementation effort for two-way transformations by using simple symmetric lenses as the initial design pattern. It ensures correctness within reason by providing a simple lens-testing framework. We demonstrate the usability of BX lenses in a realistic scenario by using \textit{Bifrons} to perform a case study experiment synchronizing data from two structurally and technologically heterogeneous databases.
\end{abstract}


\section{Introduction}

Consistent two-way transformations are a common requirement in software development. Data engineering repeatedly explores two-way data synchronization and transformation to make data ubiquitously available. Software architecture observes two-way transformations in terms of mapping between models. Even software-aided organizational design requires two-way synchronization~\cite{edrisi_developing_2024}. Data engineering tries to overcome the challenge through exchange mechanisms~\cite{jussen_issues_2024}, software design through testing~\cite{luong_system_2024}, and software-aided organizational design usually considers it a problem that falls outside its remit.

Despite their recognized importance in various subfields of software engineering, bidirectional transformations (BX) have intriguingly been an overlooked topic in the mainstream. The roots of BX research lie in the seminal work of Bancilhon and Sypratos~\cite{bancilhon_update_1981}, who introduced the practical use of BX in updatable views for databases. Their work presents the systematization of the \textit{view-update problem}, which is considered the general challenge in the field of BX. The main concern of the view-update problem is the correct preservation of changes in data made amid round-trip transformations. The reason why the results of BX research have not been applied to other subfields of software engineering is probably due to their specialized and obscure nature. While the theoretical aspect of BX is contained in the formal logic of type theory, BXs are currently performed experimentally by implementations of domain-specific languages (DSL). At the time of writing, there are several BX languages at different technological levels. The best known of these are: BiGUL~\cite{ko_bigul_2016}, HOBiT~\cite{matsuda_hobit_2018}, AUGEAS~\cite{lutterkort_augeasconguration_2008}, Boomerang~\cite{bohannon_boomerang_2007, foster_bidirectional_2010, miltner_synthesizing_2020}, and Sparcl~\cite{matsuda_sparcl_2024}.

BX languages provide a versatile environment for experimentation and testing theoretical concepts. From a technological point of view, such solutions cannot be easily combined with existing systems and code bases. A set of BXs implemented in one of the above languages must be deployed as a separate physical component and adapted to the surrounding system. In addition, the requirement for development teams to learn a new programming language to enable BX increases their cognitive load in such endeavors. These two issues are the main barriers to widespread technological adoption of BX.

Once adopted, BX can be used to construct structured and consistent two-way transformations that can be used in various aspects of software development. This paper aims to bridge the gap between the theoretical advances in BX and their practical implementation. In particular, this paper shows that BX can be:
\begin{itemize}
 \item written in a general-purpose programming language;
 \item implemented as a software library;
 \item shown to be correct within reason through unit tests;
 \item used in realistic scenarios in data engineering.
\end{itemize}

By embedding BX directly into familiar programming languages as a design pattern instead of using a DSL, the cognitive effort for developers \cite{gamma_design_1995} would be reduced and enable seamless integration of BX solutions into existing software ecosystems. As a concrete implementation of the proposed concept, this paper demonstrates a BX library in C\#. We chose C\# because it qualifies as a widely used general-purpose programming language~\cite{cass_top_2023}. The BX library is supported by a simple test framework. As an empirical proof of usability, this paper presents a proof-of-concept study in which the library is used for data synchronization between two technologically and structurally heterogeneous databases.

Section~\ref{sec:bx_preliminaries} provides a concise overview of BX with a focus on lenses. Section~\ref{sec:bifrons} deals with the implemented lens library and testing the correctness (behavedness) of the implemented lenses. Section~\ref{sec:case_study} presents an experimental case study in which data is migrated and synchronized between two relational databases using lenses from the lens library. Section~\ref{sec:conclusion} provides a summary of the research discussed in this paper.

\section{BX preliminaries}\label{sec:bx_preliminaries}

The goal of BX is to create a backward transformation from a given forward transformation. In general, BX is outlined by three common approaches~\cite{foster_three_2012}: \textit{semantic} BX~\cite{voigtlander_bidirectionalization_2009}, \textit{syntactic} BX~\cite{matsuda_bidirectionalization_2007}, and BX \textit{combinators}~\cite{foster_combinators_2007}. While the first two approaches focus on the automatic derivation of backward transformations, the BX combinator approach relies on a set of small BXs, called \textit{BX lenses}, with proven levels of correctness. The lenses are combined using \textit{combinators} to construct more complex BX lenses. Interestingly, Foster et al.~\cite{foster_three_2012} found that this approach does not allow programmers to describe lenses using programs in existing languages. As mentioned earlier, this paper intends to initiate this endeavor within reasonable bounds.

\subsection{Basic lens}
\begin{figure}[h]
    \centering
    \includegraphics[width=0.452\textwidth]{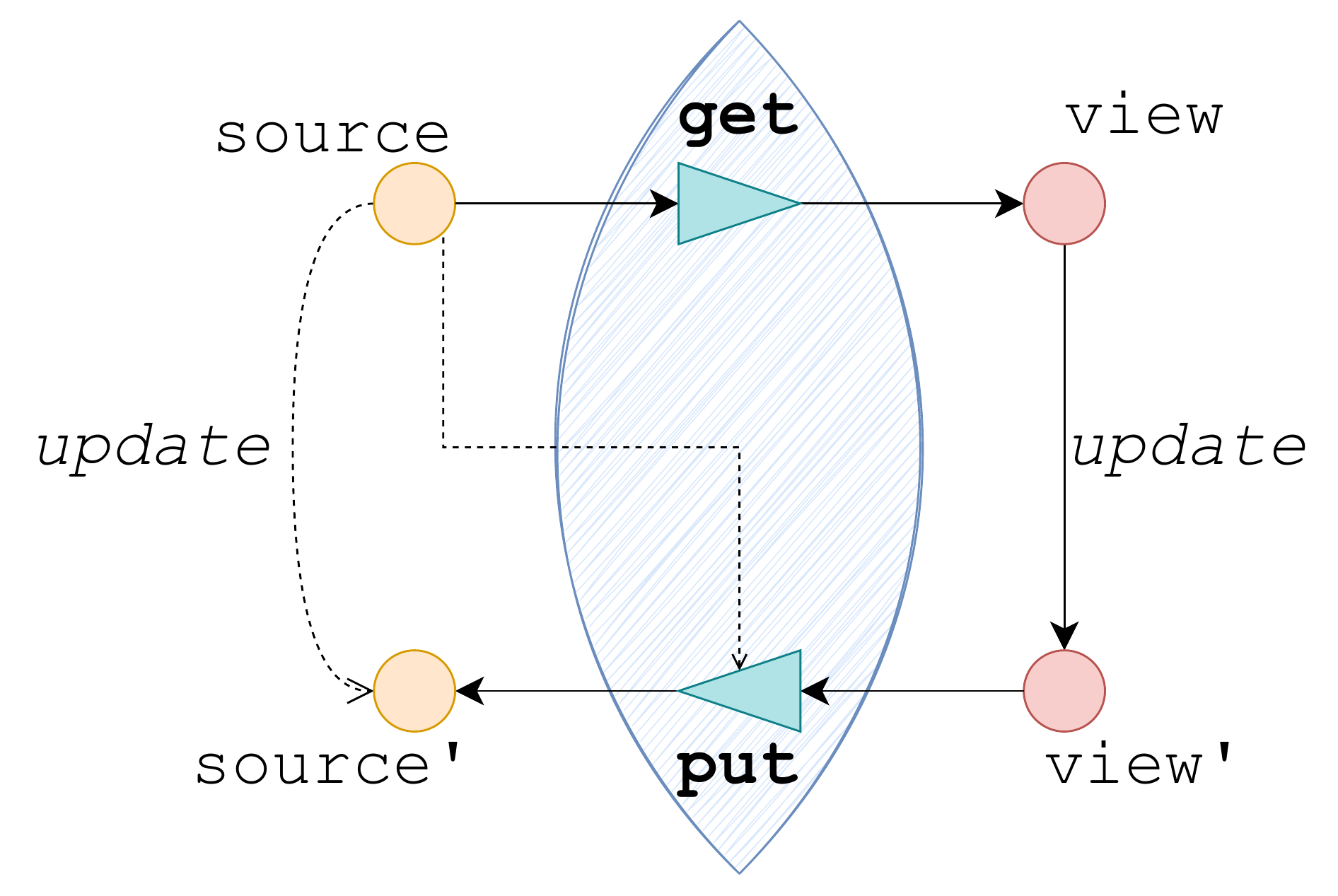}
    \caption{Basic lens}
    \label{fig:basic_lens}
\end{figure}

The most elementary structure of combinational BX is the \textit{basic lens} (Definition~\ref{def:basic_lens}). The basic lens is a pair consisting of a forward transformation function \textit{get} and a backward transformation function \textit{put}. These transformation functions operate over source and view structures, denoted as $s \in S$ and $v \in V$ respectively. The key feature of a BX lens is the ability to propagate updates on the view-structured data into the source-structured data (Figure~\ref{fig:basic_lens}).

To ensure a certain degree of transformation correctness, lenses must adhere to some of the \ref{def:putget}, \ref{def:getput}, \ref{def:puttwice} and \ref{def:putput} round-tripping laws. The more of these laws a lens obeys, the more \textit{well-behaved} it is. Behavedness is the certainty with which a lens can propagate data and its updates persistently.

Optionally, a \textit{create} function can be assigned to a lens to naively transform a view into a source without considering the original source. This addition introduces the \ref{def:createget} law.

\begin{definition}[Basic lens]\label{def:basic_lens}
Fix a universe $\mathcal{U}$ of objects and let $S \subseteq \mathcal{U}$ and $V \subseteq \mathcal{U}$ be sets of sources and views. A \textit{basic lens} $l$ from $S$ to $V$ comprises three total functions:
\begin{align*}
    l.get &\in S \rightarrow V \\
    l.put &\in V \rightarrow S \rightarrow S \\
    l.create &\in V \rightarrow S
\end{align*}
obeying the following laws for every $s \in S$ and $v \in V$:
\begin{align*}
        l.get\;(l.put\;v\;s) &= v \tag{PutGet}\label{def:putget} \\
        l.put\;(l.get\;s)\;s &= s \tag{GetPut}\label{def:getput} \\
        l.get\;(l.create\;v) &= v \tag{CreateGet}\label{def:createget}
\end{align*}
\end{definition}
A set of lenses mapping from $S$ to $V$ is written as $S \Longleftrightarrow V$; read as "put $v$ into $s$".

Stemming from the definition of a basic lens, every (well-behaved) basic lens adheres to the \ref{def:puttwice} law:
\begin{align*}
    l.put\;v\;(l.put\;v\;s) = l.put\;v\;s \tag{PutTwice}\label{def:puttwice}
\end{align*}
This law is acquired through the \ref{def:getput} and \ref{def:putget}, postulating that the $put$ should be idempotent (produce no side-effects).

The \ref{def:putput} law expresses that in a sequence of back-propagated updates, only the last update will actually affect the state of the source; updates in between should not affect the final state of the source. A lens that obeys \ref{def:putput} is called a \textit{very well-behaved} lens.
\begin{align*}
    l.put\;v_1\;(l.put\;v_2\;s)=l.put\;v_1\;s \tag{PutPut}\label{def:putput}
\end{align*}

\textit{Complements} are another mechanism that can be assigned to a lens. Complements (denoted as $c \in C$) are parts of the source that have not been mapped to a view. The complement is constructed by a function $res \in S \rightarrow C$. The complement provides additional information about the original source for the $put$ function. In this case, the $put$ function then has an alternative $l.put\;\in\;(V \times C) \rightarrow S \rightarrow S$ signature. Adjusted round-tripping laws are then applied.

If the functional interfaces on the $S$ and $V$ side of a lens differ, such lenses are called \textit{asymmetric lenses}, as in the case of the basic lens. This leads to difficulties in their composition and reuse. The mismatch of the source and view-facing interfaces may require the implementation of adapters to support composition. Asymmetric lenses only monitor updates on the $V$ side, and updates on the $S$ side are not considered or not supported, so they have a functional "sideness". Strictly speaking, for asymmetric lenses $S \Longleftrightarrow V$ is not equal to $V \Longleftrightarrow S$. This means that lenses that support updates on the opposite side must be implemented additionally, or that an inversion combinator must be used. In both cases, the use of asymmetric lenses becomes increasingly impractical.

\subsection{Simple symmetric lens}

Symmetric lenses were introduced by Hofmann et al.~\cite{hofmann_symmetric_2011} as a generalization of the asymmetric lens. Instead of considering a source and a view structure, symmetric lenses equally interface with the structures $X$ and $Y$. The sides of a symmetric lens are only labeled left and right to distinguish them semantically (e.g. $l.putr$ and $l.putl$). A notable advantage of symmetric lenses over asymmetric lenses is the ability to propagate updates on both sides of the lens. However, this comes at the cost of using complements and storing them.

\begin{figure}
    \centering
    \includegraphics[width=1\linewidth]{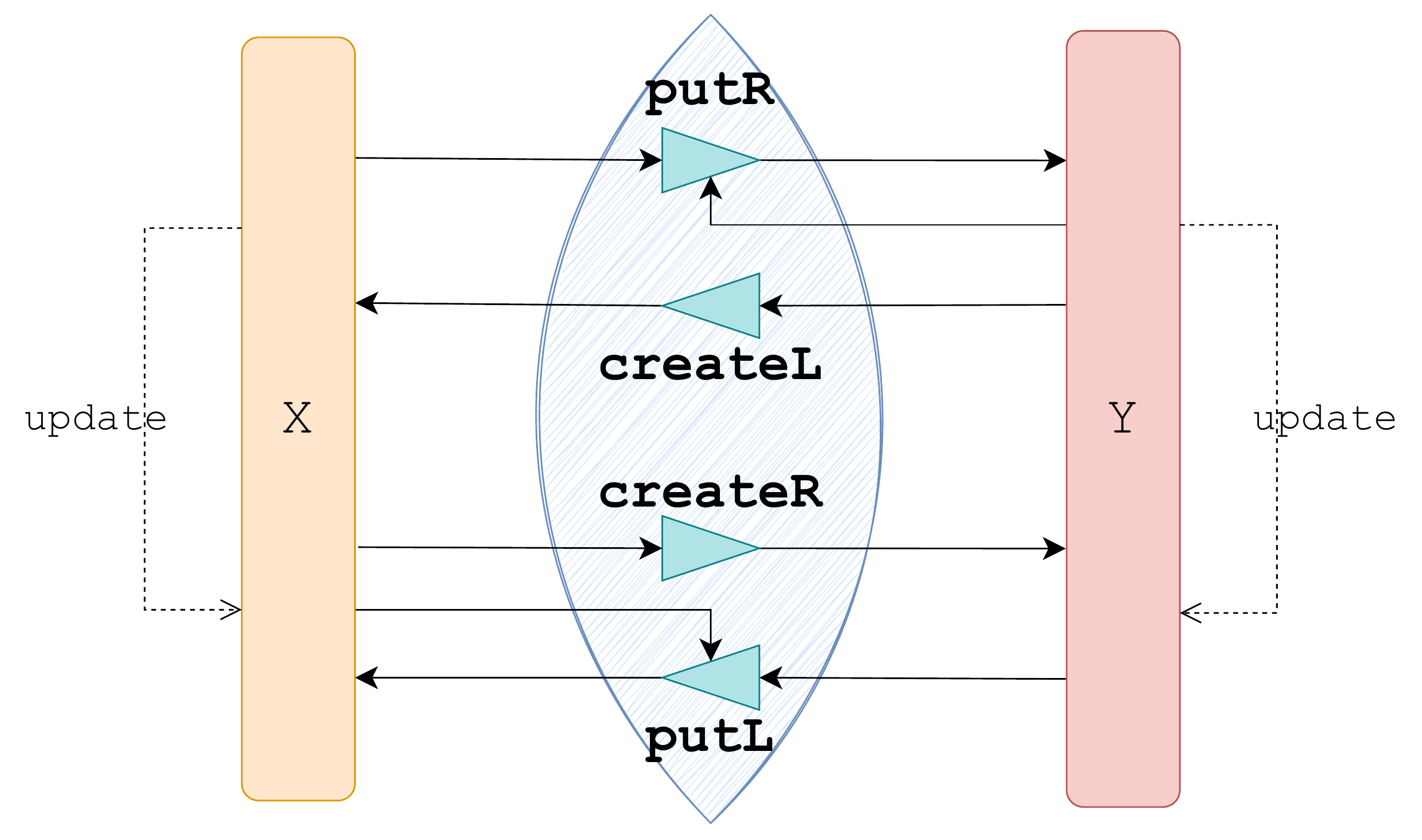}
    \caption{Simple symmetric lens}
    \label{fig:simple_sym_lens}
\end{figure}

\textit{Simple symmetric lenses} (Definition~\ref{def:simple_sym_lens}) are a further distillation of symmetric lenses that don't involve a complement \cite{miltner_synthesizing_2019,miltner_synthesizing_2020} (Figure~\ref{fig:simple_sym_lens}). 
\begin{definition}[Simple symmetric lens]\label{def:simple_sym_lens}
    Having a set of structures $X$ and $Y$, a simple symmetric lens $l \in X \Longleftrightarrow Y$ has the following components:
    \begin{align*}
        l.createR &\in X \rightarrow Y \\
        l.createL &\in Y \rightarrow X \\
        l.putR &\in X \rightarrow Y \rightarrow Y \\
        l.putL &\in Y \rightarrow X \rightarrow X
    \end{align*}
    obeying the following round-tripping laws:
    \begin{align*}
        l.putL(l.createR\;x)\;x=x \tag{CreatePutRL}\label{create_put_rl}\\
        l.putR(l.createL\;y)\;y=y \tag{CreatePutRL}\label{create_put_lr}\\
        l.putL(l.putR\;x\;y)\;x=x \tag{PutRL}\label{put_rl}\\
        l.putR(l.putL\;y\;x)\;y=y \tag{PutLR}\label{put_lr}
    \end{align*}
\end{definition}
Miltner et al.~\cite{miltner_synthesizing_2019,miltner_synthesizing_2020} showed that simple symmetric lenses are strictly more expressive than asymmetric lenses. The authors also proved that an asymmetric lens can be represented by a simple symmetric lens (Definition~\ref{def:simple_sym_asym_lens}).
\begin{definition}\label{def:simple_sym_asym_lens}
    Let $l$ be an asymmetric lens. $l$ is also a simple symmetric lens, where:
    \begin{align*}
        & l.createL\;y=l.create\;y & l.createR\;x = l.get\;x &\\
        & l.putL\;y\;x=l.put\;y\;x & l.putR\;x\;y = l.get\;x &
    \end{align*}
\end{definition}

Simple symmetric lenses were used by Miltner~\cite{miltner_synthesizing_2020} to extend the Boomerang BX language~\cite{bohannon_boomerang_2007} with lenses and combinators for strings. Atomic string transformations are provided by the lenses \textit{id}, \textit{ins} and \textit{del}. These lenses use a regex or a constant string to specify whether a part of the string should be copied, inserted, or deleted from one side to the other. The expressions in Boomerang are aligned to the left side, so a \textit{ins} lens inserts a constant on \textit{putR} and \textit{createR}, but removes it on \textit{putL} and \textit{createL}. Listing~\ref{code:boom_lenses} shows how the three elementary lenses can be combined to apply transformations between a beautified and CSV line string. Miltner~\cite{miltner_synthesizing_2020} also provided composition by multiple combinators to construct complex lenses. In particular, the combinator \textit{concat} (\texttt{'.'} operator) places lenses sequentially over a single string, as shown in Listing~\ref{code:boom_lenses}.

\begin{lstlisting}[language=Python, caption=Conceptual example of Boomerang lenses used to beautify a CSV line, label={code:boom_lenses}, morekeywords={let, ins}]
# an example csv line
let csv_line = "John;Doe;35;New York"
# elementary lenses
let l_semi = del(";")
let l_space = ins("~")
let l_comma = ins(",~")
let l_name = id("[a-zA-Z]+")
let l_nameT = ins("Name:~")
let l_age = id("\d+")
let l_ageT = ins("Age:~")
let l_city = id("[a-zA-Z~]+")
let l_cityT = ins("City:~")

let l_beautify = 
    l_nameT . l_name . l_semi . l_space . l_name . l_semi . l_comma #name
    . l_ageT . l_age . l_semi . l_comma #age
    . l_cityT . l_city #city
    
let beau = l_beautify.createR csv_line

> beau
$> "Name: John Doe, Age: 35, City: New York"
> l_beautify.createL beau
$> "John;Doe;35;New York"
\end{lstlisting}

\section{Bifrons BX lens library}\label{sec:bifrons}

We have developed a BX lens library called \textit{Bifrons} by expanding on the capabilities and ideas of Boomerang's simple symmetric lenses. The Bifrons library was written in C\# using .NET 8 to match the idea of BX in a general-purpose and widely-accepted programming language. Although Bifrons currently presents a set of prototypes, the full realisation of this achievement would make DSLs unnecessary for the implementation of BXs. We have made Bifrons fully open source to enable further experimentation and research on this topic~\cite{doncevic_bifrons_2024}.

Bifrons supports the same core lenses and combinators over strings as Boomerang, but also covers other data types and structures. Bifrons also provides a generic testing framework that can be used to check compliance with the round-tripping laws for lenses.

\subsection{Bifrons lenses}

The lenses in Bifrons are enriched with monadic results to indicate whether a transformation was successful or not. A similar case was presented by Voigtländer~\cite{voigtlander_bidirectionalization_2009} to provide partial results for BX transformations. In our case, the \texttt{Result} monad systematically denotes the outcomes of transformations. \texttt{Result} contains either an error or result data. The \texttt{Result} has an associated \texttt{bind} (infix operator "\texttt{>>=}") and \texttt{return} functions. Of course, this signature change requires adjustments to the expressions of the round-tripping laws, as shown in Definition~\ref{def:monadic_simple_symmetric_lens}.

\begin{definition}[Monad-enriched simple symmetric lens]\label{def:monadic_simple_symmetric_lens}
A monad-enriched simple symmetric lens $l \in X \Longleftrightarrow Y$ contains the following four functions:
    \begin{align*}
        createR &\in X \rightarrow \texttt{Result}\;Y \\
        createL &\in Y \rightarrow \texttt{Result}\;X \\
        putR &\in X \rightarrow Y \rightarrow \texttt{Result}\;Y \\
        putL &\in Y \rightarrow X \rightarrow \texttt{Result}\;X 
    \end{align*}
subject to four round-tripping laws expressed by:
\begin{align*}
    &createR\; x\; \texttt{>>=}\; {\lambda}y \rightarrow putL\; y\; x = \texttt{return}\; x & \tag{CreatePutRLTest} &\\
    &createL\; y\; \texttt{>>=}\; {\lambda}x \rightarrow putR\; x\; y = \texttt{return}\; y & \tag{CreatePutLRTest} &\\
    &putR\; x\; y\; \texttt{>>=}\; {\lambda}y \rightarrow putL\; y\; x = \texttt{return}\; x & \tag{PutRLTest} &\\
    &putL\; y\; x\; \texttt{>>=}\; {\lambda}x \rightarrow putR\; x\; y = \texttt{return}\; y & \tag{PutLRTest} &
\end{align*}
\end{definition}

Listing~\ref{code:bifrons_string_lenses} offers the Boomerang example from Listing~\ref{code:boom_lenses}, but translated into Bifrons. Despite the obvious syntax change to adhere to the C\# coding style, the semantics of the code remain the same\footnote{This and the following examples in this section can be found in the Experiments project in the file \texttt{Bifrons.Experiments/PaperExamples.cs}}. Instead of the usual \texttt{'.'} lens concatenation operator, C\# operator overloading restricts the selection to the use of \texttt{'\&'} as an equivalent operator.

\begin{lstlisting}[language=CSharp, caption=Implementation of lenses from Listing~\ref{code:boom_lenses} using the Bifrons library, label={code:bifrons_string_lenses}]
var csvLine = "John;Doe;35;New York";
var exp_csvLine = "John;Doe;35;New York";
var exp_beautified = "Name: John Doe, Age: 35, City: New York";

var l_semi = %DeleteLens%.Cons(";");
var l_space = %InsertLens%.Cons(" ");
var l_comma = %InsertLens%.Cons(", ");
var l_name = %IdentityLens%.Cons(@"[a-zA-Z]+");
var l_nameT = %InsertLens%.Cons("Name: ");
var l_age = %IdentityLens%.Cons(@"\d+");
var l_ageT = %InsertLens%.Cons("Age: ");
var l_city = %IdentityLens%.Cons(@"[a-zA-Z ]+");
var l_cityT = %InsertLens%.Cons("City: ");

var l_beautify = l_nameT & l_name & l_semi & l_space & l_name & l_semi & l_comma  // name
    & l_ageT & l_age & l_semi & l_comma // age
    & l_cityT & l_city; // city

var res_beautified = l_beautify.CreateRight(csvLine);
%Assert%.True(res_beautified);
var beautified = res_beautified.Data;
%Assert%.Equal(exp_beautified, beautified);

var res_csvLine = l_beautify.CreateLeft(beautified);
%Assert%.True(res_csvLine);
var resultingCsvLine = res_csvLine.Data;
%Assert%.Equal(exp_csvLine, resultingCsvLine);
\end{lstlisting}

Although string lenses provide convenient examples, a versatile lens library should also contain other types to be practical. Bifrons additionally contains lenses for the boolean, date-time, decimal, integer, and long data types. The combinators for each type have been adapted to their respective algebras. The main difference is that strings are essentially aggregate types. Listing~\ref{code:bifrons_other_type_lenses} demonstrates the use of addition and subtraction lenses for integers combined by a general-purpose sequential composition lens combinator initialized by the operator "\texttt{>>}". The listing also contains a cross-type lens that consistently maps between integer and string.

\begin{lstlisting}[language=CSharp, caption=Example of Bifrons integer and cross-type lenses with combinators, label={code:bifrons_other_type_lenses}]
var l_id = %IdentityLens%.Cons();
var l_add1 = %AddLens%.Cons(1);
var l_add5 = %AddLens%.Cons(5);
var l_sub3 = %SubLens%.Cons(3);

var l_comp = l_id >> l_add1 >> l_add5 >> l_sub3;

int left = 10;
int expectedRight = 13;
int expectedUpdatedLeft = 8;

var right = l_comp.CreateRight(left); // 10+(1+5-3)=1

var updatedRight = right.Data - 2; //13-2=11 
var updatedLeft = l_comp.PutLeft(updatedRight, left); // (-1-5+3)+11=8

// CrossType example
var l_intStr = %CrossType%.%IntegerStringLens%.Cons();
var l_compStr = %Combinators%.Compose(l_comp, l_intStr); // w/o operator
var rightString = l_compStr.CreateRight(left); // 10+(1+5-3)=13=>"13"
\end{lstlisting}

Listing~\ref{code:bifrons_combinators} shows an example of an advanced \texttt{Or} combinator that works with the monadic \texttt{Either} type. This combinator was proposed by Foster~\cite{foster_combinators_2007} and Bohannon et al.~\cite{bohannon_boomerang_2007} in a monomorphic environment (which only supports strings). In Bifrons, this lens is polymorphic and supports different data types on both sides of the \texttt{Either} monad. This is useful for managing type-heterogeneous data streams, such as those that can occur with unstructured data sources. Listing~\ref{code:bifrons_combinators} is a toy problem solution for a combined lens that creates anonymized monthly birthday reminders. In the example, the incoming data is accepted as either \texttt{DateTime} or \texttt{string}, the year of birth is removed along with the redundant month, and the remaining day information is subtracted by one to set the reminder to a day earlier\footnote{This naive example will produce an erroneous 0 if a person's birthday falls on the 1\textsuperscript{st} of a month}.

\begin{lstlisting}[language=CSharp, caption=Example of more advanced lens combinations from Bifrons, label={code:bifrons_combinators}]
// (str<=>datetime)>>(datetime<=>int)>>(int<=>int)
var l_ifStr = %Combinators%.Compose( 
    %Combinators%.Compose(
        %CrossType%.%StringDateTimeLens%.Cons(),
        %DateTimes%.%DayLens%.Cons()
    ),
    %Integers%.%SubLens%.Cons(1)
);
// (datetime<=>int)>>(int<=>int)
var l_ifDateTime = %Combinators%.Compose( 
    %DateTimes%.%DayLens%.Cons(),
    %Integers%.%SubLens%.Cons(1)
);
// (str<=>datetime)>>(datetime<=>int)>>(int<=>int) 
//  or (datetime<=>int)>>(int<=>int)
var l_or = %Combinators%.Or(l_ifStr, l_ifDateTime);

string dateTimeString = %Either%.Left<string, DateTime>(new DateTime(1992, 12, 31).ToString("yyyy-MM-ddTHH:mm:ss"));
%DateTime% dateTime = %Either%.Right<string, DateTime>(new DateTime(1992, 12, 31)); 

var res_fromString = l_or.CreateRight(dateTimeString); // returns 30: int
var res_fromDateTime = l_or.CreateRight(dateTime); // returns 30: int

var res_returnOriginalDateString = l_or.PutLeft(
    %Either%.Left<int, int>(res_fromString.Data.Left + 1),
    dateTimeString
    ); // returns Left("1992-12-31T00:00:00")
\end{lstlisting}

\subsection{Testing the lenses}
The usefulness of BX lenses lies in the round-tripping laws. Lens interfaces from the Bifrons library can be implemented arbitrarily, and the interfaces alone don't guarantee that the implementation contains the correct transformations. For each lens implemented, correctness must be proven to a level of behavedness. The concept of proving the behavedness of lenses has so far been done by formal logical and mathematical proofs. This is possible in functional languages such as Haskell or OCaml, as a series of expression substitutions can lead to a proof. Bifrons was implemented in C\#, which is not primarily a functional language and has a rich ecosystem of expressions for constructing programs. This makes formal proofs for Bifrons lenses difficult and exhaustive. C\# expressions used in implementations can be constrained by the use of complex type constructs or reflection, making the implementation of lenses inherently more complex than any other piece of code. This would be a setback to our original goal of seamlessly integrating lenses into a widely used general-purpose programming language.

The Bifrons library offers a more practical solution for proving the behavedness of lenses - through empirical proof via unit tests. The empirical proof is not as comprehensive as formal proof, but it makes it possible to prove the correctness of lenses within reasonable bounds. Designing and writing tests is a typical task for software engineers. Most software relies on the reasoning of its developers to show that it works or is correct within reason. When implementing Bifrons, we assumed that in industry environments reasonable unit tests written by the developers are sufficient to show the behavedness of lenses. With this mindset, we have created a way to explore the trade-off between formalization and feasibility in BX lenses. We believe that this is the main obstacle to wider adoption of BX lenses.

Developers using Bifrons' lenses do not need to worry about lens or combinator testing, but if they have implemented a new lens from scratch, appropriate testing is expected. Developers are guided through the test implementation by the small testing framework within the Bifrons library. The tests representing the round-tripping laws are generically implemented in a partially implemented abstract class (Listing~\ref{code:bifrons_test_fw}). Each test method represents one of the round-tripping laws for simple symmetric lenses from Definition~\ref{def:monadic_simple_symmetric_lens}: \ref{create_put_rl}, \ref{create_put_lr}, \ref{put_rl}, and \ref{put_lr}. Tests analogous to \ref{def:putput} are also included to check the correctness of the update propagation from both sides of the lens \footnote{The entire testing class can be found in the project repository at \texttt{/Bifrons.Lenses.Tests/SymmetricLensTesting.cs}}.

\begin{lstlisting}[language=CSharp, caption=Tests for lens laws as found in the Bifrons lens testing framework, label={code:bifrons_test_fw}]

public override void CreatePutLRTest()
{
    var result =
    _lens.CreateLeft(_right)
        .Bind(left => _lens.PutRight(left, %Option%.Some(_right)));

    %Assert%.True(result);
    %Assert%.Equal(_right, result.Data);
}

public override void CreatePutRLTest()
{
    var result =
    _lens.CreateRight(_left)
        .Bind(right => _lens.PutLeft(right, %Option%.Some(_left)));

    %Assert%.True(result);
    %Assert%.Equal(_left, result.Data);
}

public override void PutLRTest()
{
    var result =
    _lens.PutLeft(_right, %Option%.Some(_left))
        .Bind(left => _lens.PutRight(left, %Option%.Some(_right)));

    %Assert%.True(result);
    %Assert%.Equal(_right, result.Data);
}

public override void PutRLTest()
{
    var result =
    _lens.PutRight(_left, %Option%.Some(_right))
        .Bind(right => _lens.PutLeft(right, %Option%.Some(_left)));
        
    %Assert%.True(result);
    %Assert%.Equal(_left, result.Data);
}
\end{lstlisting}

When implementing tests for lenses, the partially implemented abstract framework class only requires a lens developer to specify:
\begin{enumerate}
 \item The default left data;
 \item The default right data;
 \item The lens under test;
 \item The expected round-tripping data for updates on the left side;
 \item The expected round-tripping data for updates on the right side
\end{enumerate}
The correctness of the tests naturally depends on the representative quality of the data provided by the developer. This is similar to testing any software code. Listing~\ref{code:bifrons_id_tests} provides an example of a test specification for a string identity lens.

\begin{lstlisting}[language=CSharp, caption=String identity lens tests specification (implementation points marked in comments), label={code:bifrons_id_tests}]
public class %IdentityLensTests% : %SymmetricLensTestingFramework%<string, string>
{
    protected override string _left => "Hello, World!"; // 1.

    protected override string _right => "Hello, World!"; // 2.

    private readonly string _greetingRegex = @"Hello, [a-zA-Z]+!";
    protected override ^ISymmetricLens^<string, string> _lens => %IdentityLens%.Cons(_greetingRegex); // 3.

    protected override (string originalSource, string expectedOriginalTarget, string updatedTarget, string expectedUpdatedSource) _roundTripWithRightSideUpdateData
        => ("Hello, World!", "Hello, World!", "Hello, Universe!", "Hello, Universe!"); // 4.

    protected override (string originalSource, string expectedOriginalTarget, string updatedTarget, string expectedUpdatedSource) _roundTripWithLeftSideUpdateData
        => ("Hello, World!", "Hello, World!", "Hello, Universe!", "Hello, Universe!"); // 5.
}
\end{lstlisting}

\subsection{BX lenses on structures and structured data}
Primitive types are only the basis for further development of lenses on aggregate and tree-like data types; we will refer to these types as \textit{data structures}. By supporting data structures, BX lenses can be used in a larger and more realistic number of use cases. While primitive types are atomic and have a built-in structure within the programming language itself, handling transformations of structured data depends on the coupling of the structural specification and the data within the structure itself. This is a challenge in the context of BX lenses.

We have approached this challenge by breaking it down into two problems:
\begin{itemize}
 \item BX transformation of data structures by BX lenses;
 \item BX transformation of data coupled with the data structures.
\end{itemize}
This approach produces structural transformations without coupling them with data transformations. This makes the lenses over structures reusable for structural mapping. The use of lenses over structures has already been investigated in detail by Diskin et al.~\cite{diskin_algebraic_2008, diskin_state-_2011, diskin_compositionality_2017} with regard to the synchronization of heterogeneous data models. The structural lenses implemented in Bifrons via this approach can be used as a new driver for empirical research and the development of lenses for data modeling.

We strategize that the BX transformations of data structures are covered by \textit{structural lenses} and the BX transformations of structured data are covered by \textit{structured data lenses}. In Bifrons, these two kinds of BX lenses aren't explicitly specified, but we give an example of structural and structured data lenses for a relational database.

\subsubsection{Structural BX lenses}
\begin{figure}[h]
    \centering
    \includegraphics[width=1\linewidth]{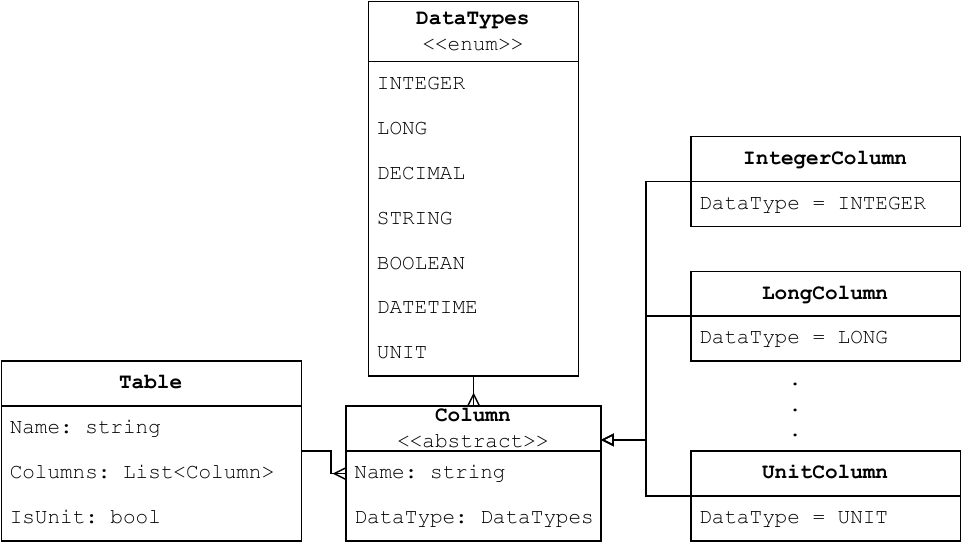}
    \caption{Relational metamodel for the Bifrons relational structural lenses}
    \label{fig:relational_metamodel}
\end{figure}

Bifrons contains a simplified metamodel describing a relational schema and a set of BX lenses within that metamodel: $Relational \Longleftrightarrow Relational$. The metamodel contains classes that describe tables and columns of each supported data type, where a unit column is used to denote a deleted column in a table. Similarly, data of the \texttt{UNIT} type is considered deleted. The metamodel is conceptually illustrated by the diagram in Figure~\ref{fig:relational_metamodel}.

\begin{figure*}[h]
    \centering
    \includegraphics[width=0.8\linewidth]{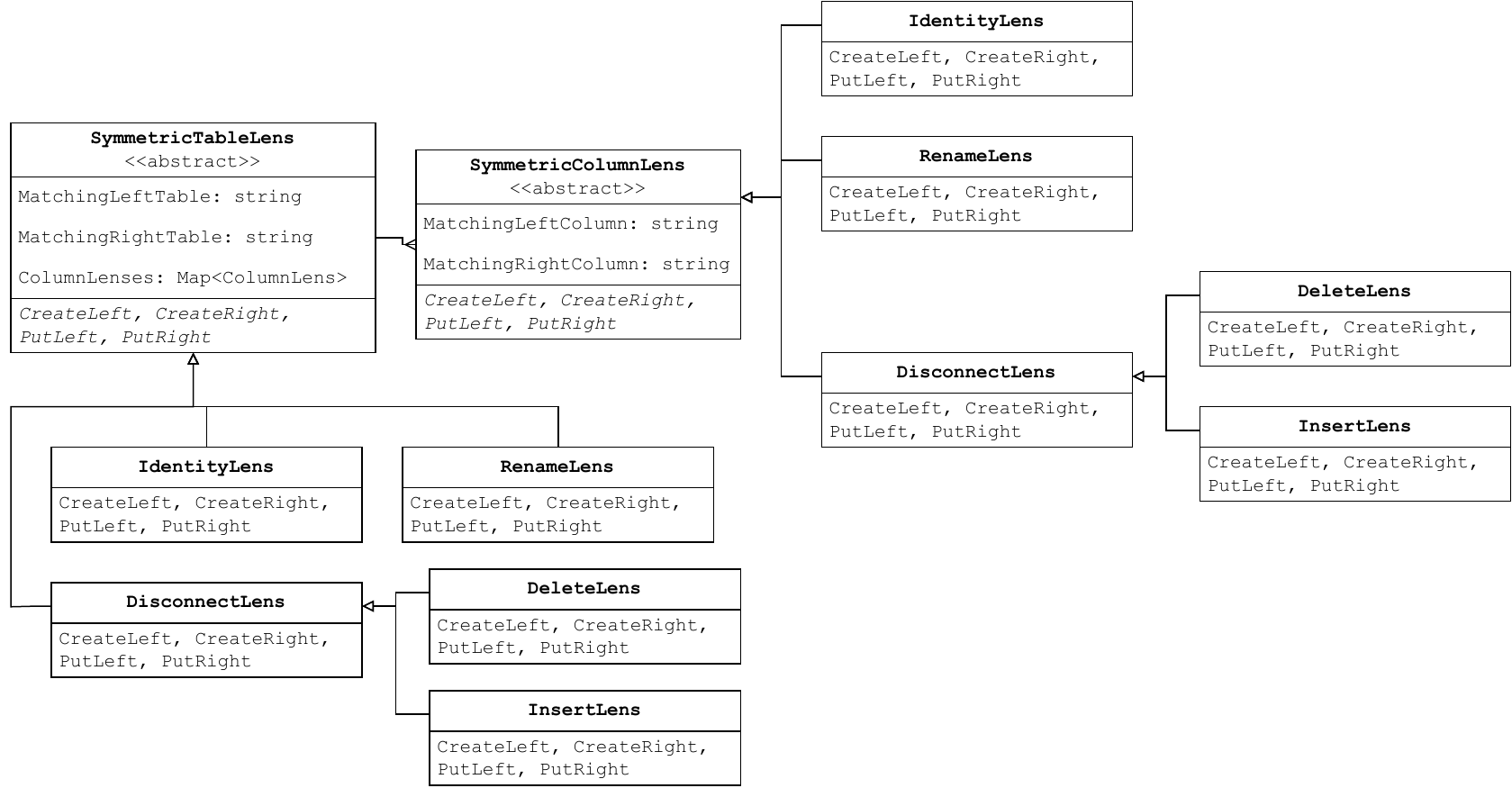}
    \caption{Relational structural BX lenses in Bifrons}
    \label{fig:relational_lenses}
\end{figure*}

The \textit{relational (structural) lenses} in Bifrons are implemented according to the same copy, delete and insert pattern as data lenses (Figure~\ref{fig:relational_lenses}). This pattern applies to both tables and columns. Interestingly, Bifrons proves here empirically that the postulate that the \texttt{disconnect} lens is the basis for the lenses \texttt{insert} and \texttt{delete}~\cite{bohannon_boomerang_2007,miltner_synthesizing_2020} also applies to the relational model. Bifrons also introduces a \texttt{rename} lens, which can be used to rename the model element while copying.

Listing~\ref{code:bifrons_relational_lenses} provides an example of the construction of a lens for a table identity lens, which copies a \texttt{People} between metamodels. The table is copied trivially, but its columns are specified to either be copied, renamed, inserted, or deleted according to a list of individual column lenses. The table on the left is set to contain columns for the identifier, first name, last name, date of birth, address, and phone number, while the table on the right is set to contain columns for the identifier, first name, last name, renamed date of birth column, and email. In general, the \texttt{identity} and \texttt{rename} lenses indicate which elements are on both sides of the lens, the \texttt{delete} lenses indicate which elements only exist on the left side, and the \texttt{insert} lenses indicate which elements only exist on the right side. It is also important to note that although the table identity lens is a lens, it's also a column lens combinator because it aggregates the effects of column lenses.

\begin{lstlisting}[language=CSharp, caption=Example of a BX table lens construction for a \texttt{People} table, label={code:bifrons_relational_lenses}]
var l_col_id = %Columns%.%IdentityLens%.Cons("Id");
var l_col_firstName = %Columns%.%IdentityLens%.Cons("FirstName");
var l_col_lastName = %Columns%.%IdentityLens%.Cons("LastName");
var l_col_dateOfBirth = %Columns%.%RenameLens%.Cons("DateOfBirth", "DoB");
var l_col_address = %Columns%.%DeleteLens%.Cons("Address");
var l_col_email = %Columns%.%InsertLens%.Cons("Email");
var l_col_phone = %Columns%.%DeleteLens%.Cons("Phone");
var l_tbl_people = 
    %Tables%.%IdentityLens%.Cons(
        "People", col_id, col_firstName, 
        col_lastName, col_dateOfBirth, col_address, 
        col_email, col_phone);
\end{lstlisting}

\subsubsection{Structured data lenses}
The key concept of structured data lenses is that they contain a pair of structural and data lenses. The structural lenses are used to align the data structure with the structure specification on both sides of the lenses. References to structural lenses can be assigned redundantly in the structured data lenses to calculate structural snapshots on the fly. The structured data lenses use data lenses to transform the concrete data within the structure itself. As with the data and structural lenses, the structured data lenses can also be combined to form more complex lenses. The combination of structured data lenses brings additional complexity, as both the structural and the data mapping must be specified.

\begin{figure*}[h]
    \centering
    \includegraphics[width=0.8\linewidth]{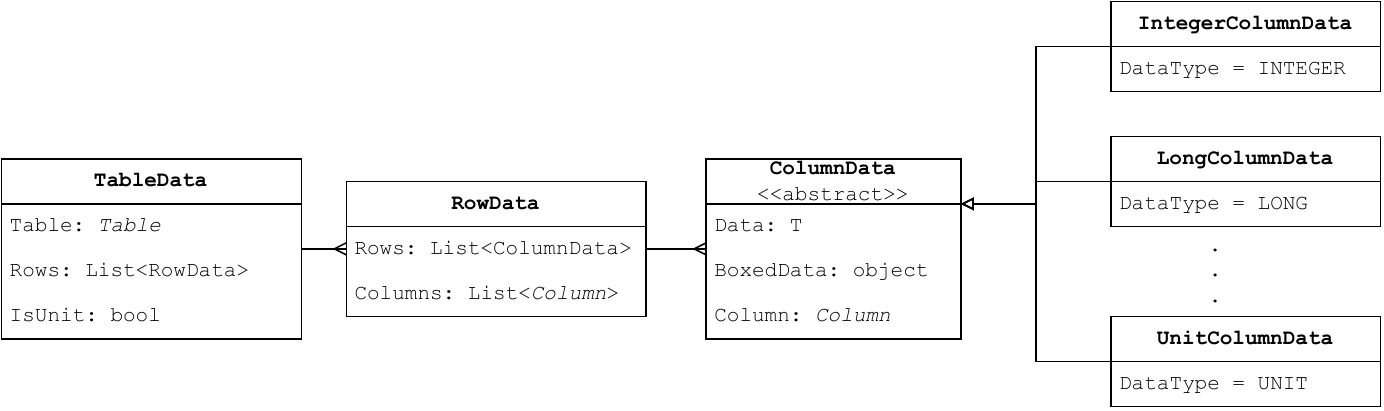}
    \caption{Relational data model in Bifrons (referenced relational model elements are omitted)}
    \label{fig:relational_data_model}
\end{figure*}

As an example of structured data lenses, Bifrons currently contains structured data lenses for relational data models. The \textit{relational data lenses} use relational lenses in conjunction with data lenses for primitive types to support the BX transformations of relational data instances. In Bifrons, a data-enriched relational model was developed to describe both a relational model and its data (Figure~\ref{fig:relational_data_model} describes this model conceptually).

\begin{figure*}[h]
    \centering
    \includegraphics[width=0.8\linewidth]{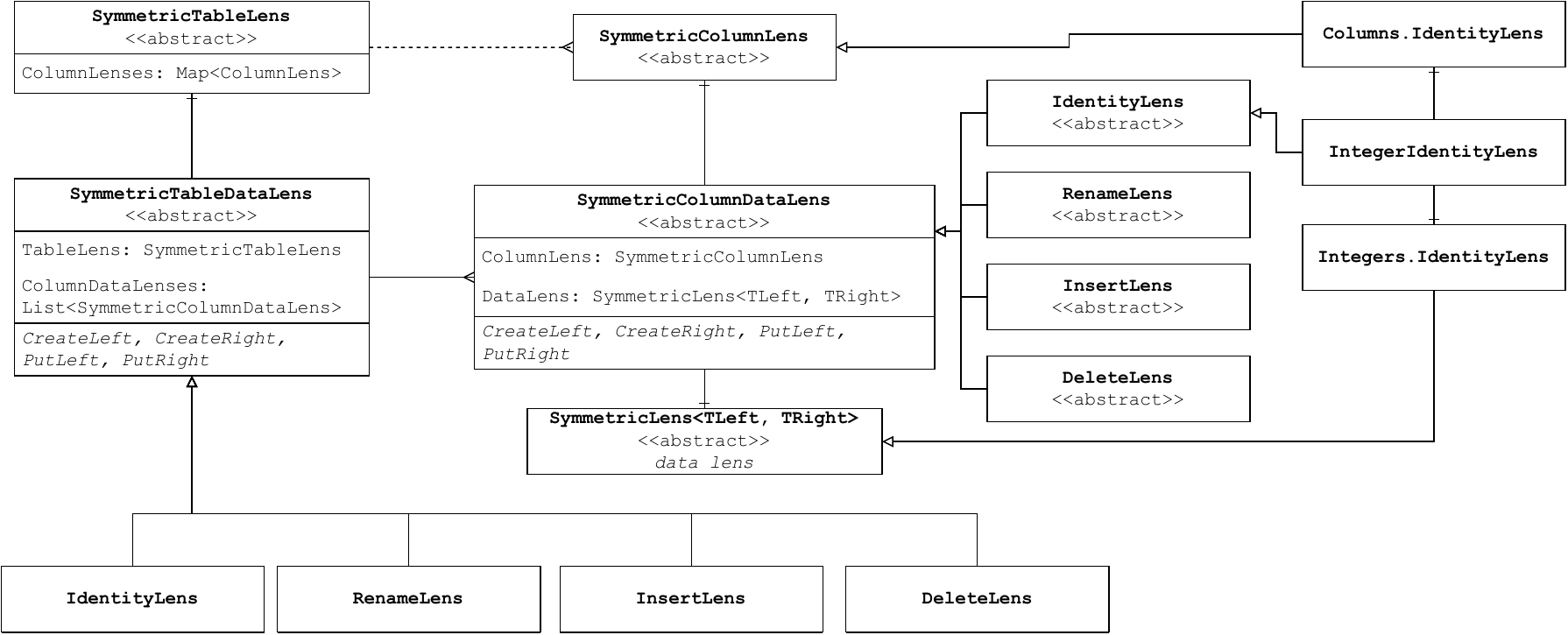}
    \caption{Relational data lenses in Bifrons}
    \label{fig:relational_data_lenses}
\end{figure*}

Relational data lenses in Bifrons offer transformations over the relational data model (Figure~\ref{fig:relational_data_lenses}). The operational capabilities still include the \texttt{identity}, \texttt{rename}, \texttt{insert}, and \texttt{delete} lenses. Listing~\ref{code:bifrons_relational_data_lenses} demonstrates the combination of a table data lens that copies data for a new \texttt{People} table.

\begin{lstlisting}[language=CSharp, caption=Example of a BX table data lens construction for a \texttt{People} \\table, label={code:bifrons_relational_data_lenses}]
// prepare column structural lenses
var l_col_id = %Columns%.%IdentityLens%.Cons("Id");
var l_col_name = %Columns%.%IdentityLens%.Cons("Name");
var l_col_dob = %Columns%.%RenameLens%.Cons("DOB", "DateOfBirth");
var l_col_isAdmin = %Columns%.%DeleteLens%.Cons("IsAdmin");
var l_col_hoursClocked = %Columns%.%InsertLens%.Cons("HoursClocked", %RelationalModel%.%DataTypes%.DECIMAL);
// prepare table structural lens
var l_tbl_people = %Tables%.%IdentityLens%.Cons(
    "People",
    [l_col_id, l_col_name, l_col_dob, l_col_isAdmin, l_col_hoursClocked]);
// prepare column data lenses
var dl_id = %DataColumns%.%IntegerIdentityLens%.Cons(
    l_col_id, %Integers%.%IdentityLens%.Cons());
var dl_name = %DataColumns%.%StringIdentityLens%.Cons(
    l_col_name, %Strings%.%IdentityLens%.Cons());
var dl_dob = %DataColumns%.%DateTimeIdentityLens%.Cons(
    l_col_dob, %DateTimes%.%IdentityLens%.Cons());
var dl_isAdmin = %DataColumns%.%BooleanDeleteLens%.Cons(
    l_col_isAdmin, false);
var dl_hoursClocked = %DataColumns%.%DecimalInsertLens%.Cons(
    l_col_hoursClocked, 0.0);
// prepare table data lenses
var dl_tbl_people = %DataTables%.%IdentityLens%.Cons(
    l_tbl_people,
    [dl_id, dl_name, dl_dob, dl_isAdmin, dl_hoursClocked])
    .Match(_ => _, msg => throw new %Exception%(msg));
\end{lstlisting}

It is important to note that not all structural and data lenses can be combined to form a structured data lens. The combined lenses must be compatible. To illustrate this problem - one might specify a structured data lens with a \texttt{delete} structural lens and an \texttt{insert} data lens. In this case, when transforming to the right, the structural \texttt{delete} lens should overpower the \texttt{insert} lens, as no data can exist if the structure (column) itself doesn't exist on the right side. The transformation to the left is problematic because the \texttt{delete} lens restores the column structurally, but the \texttt{insert} lens ignores all previous data on the left side and puts the data in a \texttt{UNIT} state (the \texttt{insert} lens assumes that there is no data at all on the left side). The Bifrons library doesn't allow such specifications in \texttt{insert} or \texttt{delete} structured data lenses. In Bifrons, the lens types for structural lenses are strictly defined and analogue data lenses are fixed by definition. This means that the developer can only set the default values for the lenses.

\subsection{On canonizers}

Canonizers are software components that translate data from one format to another, where one of these formats is considered canonical. We consider data translation as data transformation from one metamodel to another. In terms of BX lenses, canonizers were used to standardize text~\cite{foster_quotient_2008,maina_synthesizing_2018} before administering transformations. We propose extending the method of canonizers to support metamodel heterogeneity.

We have decided to interface Bifrons lenses with other data formats by using canonizers to introduce lenses in realistic scenarios. The Bifrons lenses work via C\# objects, which makes them the canonical format. C\# objects need to be translated into formats that can be used outside lens-driven programs. We have demonstrated this by applying canonizers for PostgreSQL and MySQL databases in Section~\ref{sec:case_study}.

\section{Case study experiment - relational database synchronization}\label{sec:case_study}
\begin{figure}[h]
    \centering
    \includegraphics[width=\linewidth]{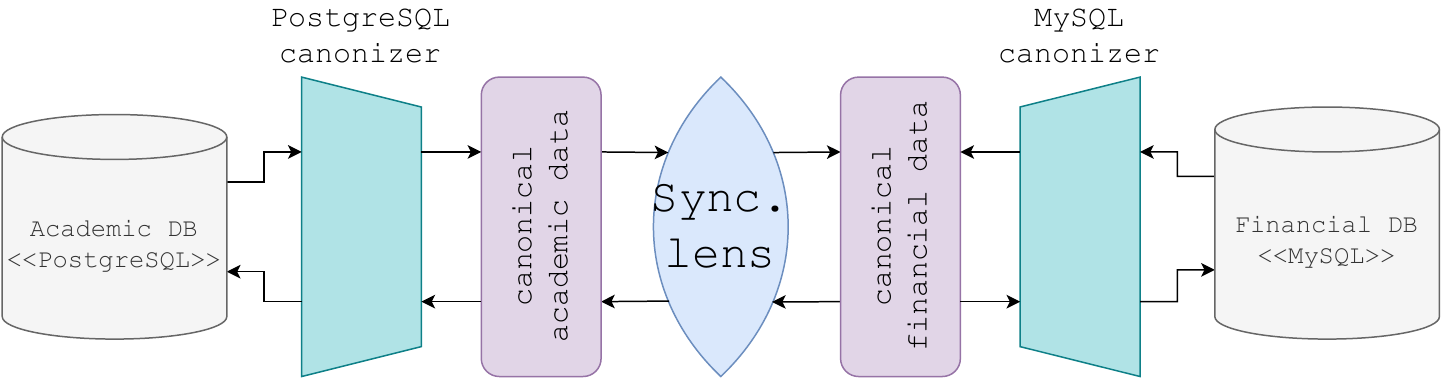}
    \caption{Lens with canonizers for PostgreSQL and MySQL data}
    \label{fig:lens_with_db_canonizers}
\end{figure}

This case study presents a use case in data engineering where data needs to be synchronized between two heterogeneous data sources. We use two canonizers to translate data to and from the canonical form, and we use a combined structured data lens to support data synchronization (Figure~\ref{fig:lens_with_db_canonizers}) \footnote{The tests that make up the case study can be found in the project repository at \texttt{/Bifrons/Bifrons.Experiments/\\RelationalDataSyncExperiments.cs}. The tests are configurable to support the change of database vendor in both domains}.

We have chosen a scenario of a fictitious university that manages academic data in a PostgreSQL database and financial data in a MySQL database. This presents a difficult case of data synchronization between structurally and technologically heterogeneous databases. The two databases semantically share a \texttt{Students} table, albeit with different and mutually confidential columns. The data of the shared columns should be synchronized between the tables, but exclusive data should be masked or omitted.

\begin{figure}[h]
    \centering
    \includegraphics[width=\linewidth]{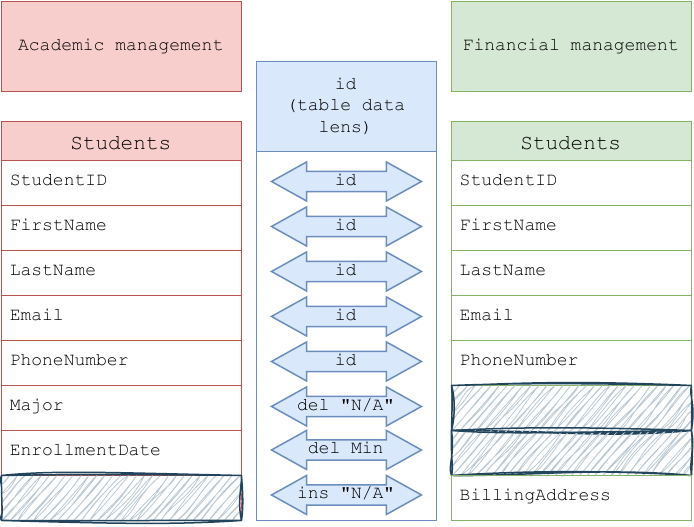}
    \caption{\texttt{Students} table mapping via relational lenses}
    \label{fig:students_table_lensing}
\end{figure}

As shown conceptually in Figure~\ref{fig:students_table_lensing}, the table data identity lens is the synchronization lens. It is combined using column data lenses. Identity lenses are used for the common columns of the tables. The columns \texttt{Major} and \texttt{EnrollmentDate} are transformed by the delete lens which removes them from the financial management's table. \texttt{BillingAddress} is transformed by the insert lens, which leaves the data in the financial management's table or sets it to a default value and removes it from the academic management's table.

Data can be naively synchronized with the \texttt{Create} function, which ignores all existing data on the other side. This is useful for instantiating a database from an existing database, i.e. for migrating to another database instance or another database model.

It can be generally assumed that both sides are filled with data and that the naive approach wouldn't be sufficient to synchronize the databases, as it would completely delete the existing data. The \texttt{Put} functions can initiate a synchronization that takes the existing data into account. In the case where the \texttt{Students} table of the financial management is updated with the academic management's data, \texttt{PutRight} must be called by passing the current academic data as the left source, followed by the current financial data as the right view.

If a complete synchronization of both sides is required, then the \texttt{PutRight} and \texttt{PutLeft} must be called in composition. This will first update the financial database and then the academic database.

The case study experiment is designed as a unit test. Docker containers are used to handle databases for financial and academic management. The choice of database vendor can be changed via a configuration file in which the container image to be created and the instantiated canonizer are marked. Once the containers are prepared to run the tests, the canonizers extract and canonize the data from the databases, the synchronization lens applies the transformations, and the canonizers store the modified data in their respective databases.

The full test run of the experiment includes an overlapping set of \texttt{Students} rows and two disjoint sets in each database. The tests assert that the data in both databases is the same modulo the lens transformations. This can also be checked manually with an SQL client.

\begin{figure}[h]
    \centering
    \includegraphics[width=\linewidth]{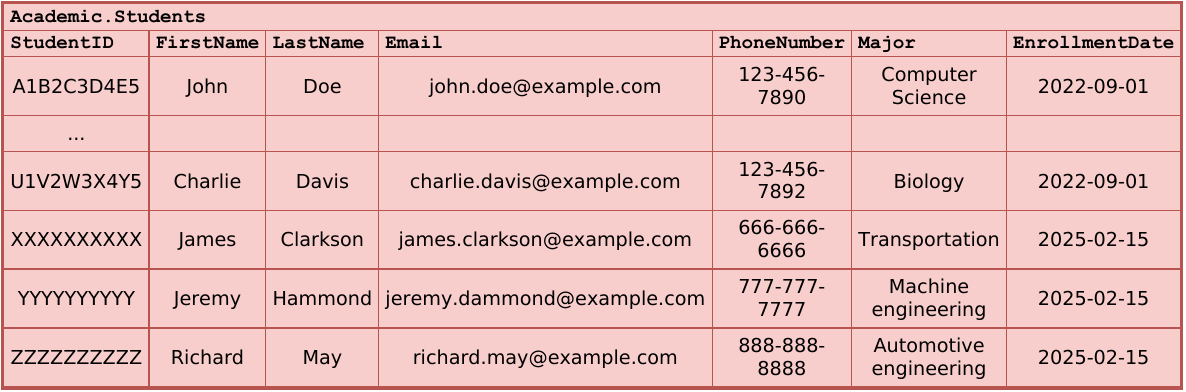}
    \caption{The academic \texttt{Students} table before synchronization}
    \label{fig:exp_before_academic}
\end{figure}

\begin{figure}[h]
    \centering
    \includegraphics[width=\linewidth]{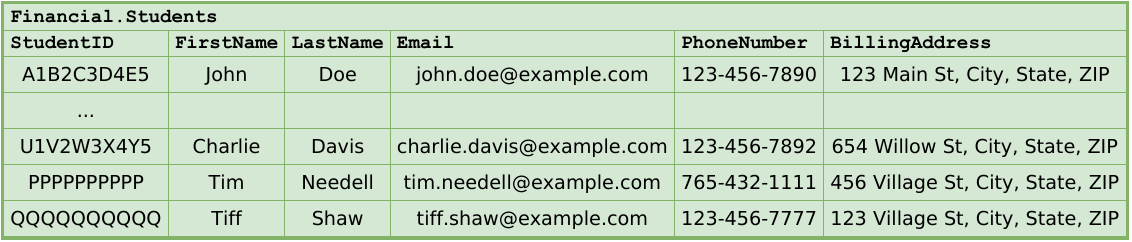}
    \caption{The financial \texttt{Students} table before synchronization}
    \label{fig:exp_before_financial}
\end{figure}

\begin{figure}[H]
    \centering
    \includegraphics[width=\linewidth]{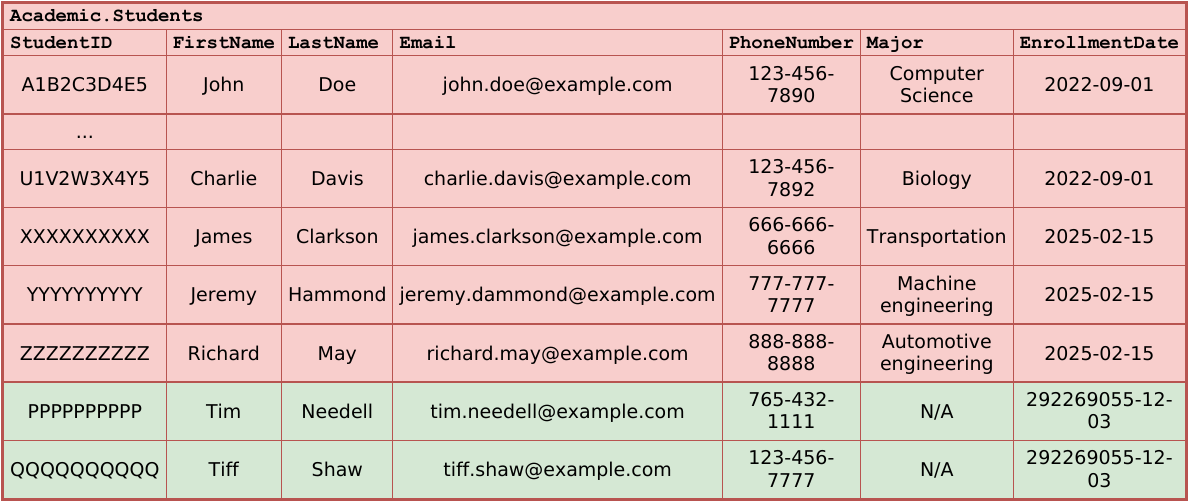}
    \caption{The academic \texttt{Students} table after synchronization (green rows denote tuples taken from the financial database)}
    \label{fig:exp_after_academic}
\end{figure}

\begin{figure}[H]
    \centering
    \includegraphics[width=\linewidth]{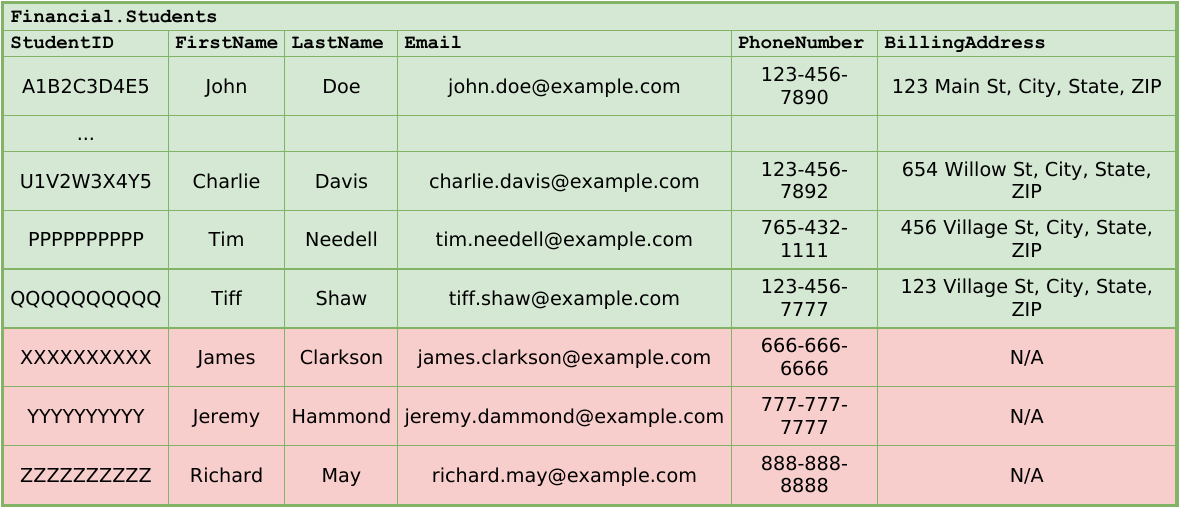}
    \caption{The financial \texttt{Students} table after synchronization (red rows denote tuples taken from the academic database)}
    \label{fig:exp_after_financial}
\end{figure}

Figures~\ref{fig:exp_before_academic} to~\ref{fig:exp_after_financial} show the data extracts before and after synchronization. Figure~\ref{fig:exp_after_financial} shows the status of the financial \texttt{Students} table after synchronization, with the red rows coming from the academic management database. Similarly, the green rows in Figure~\ref{fig:exp_after_academic} originate from the financial management database.

With this case study experiment we have shown that BX lenses can be used in a realistic scenario. As long as the proper lenses are implemented, tested, and combined, with the inclusion of canonizers, data translation can be created in a structured, manageable, and proven way. Most importantly, from a developer's perspective, the case study experiment demonstrates a BX solution that is technologically simpler than using a BX DSL and integrating it as a sidecar. There was no need to move any part of the solution outside the C\# .NET ecosystem. This lowered the technical complexity of the task at hand and consequently reduced the cognitive load on the developers.

\section{Conclusion}\label{sec:conclusion}

The introduction of BX lenses is the key to achieving structured and consistent data transformations in codebases. This inclusion can only happen through broad acceptance in software development. We have proposed implementing BX lenses in C\# to promote their adoption by developers and engineers. To prove this concept, we have implemented our own BX lens library - \textit{Bifrons}. With Bifrons, we have provided the means to construct data, structural, and structured data lenses.

By using C\#, we are limited in terms of incomplete formal proofs. To address this challenge, we have made it possible to assert the behavior of lenses within practical reasoning through unit tests. To reduce the cognitive effort for potential developers, we have provided a framework for testing lenses as a guide.

By adding canonizers to the implemented lenses, we have enabled the lenses to act on software entities outside the bidirectional program itself. Our case study experiment has shown how a combined lens can be used to synchronize data between structurally and technically heterogeneous databases. In this way, we have created a new practical solution for data synchronization in data engineering. Additionally, our solution provided a technologically simpler alternative to achieving BX than through a DSL. The solution was kept within the C\# .NET programming ecosystem, avoiding the need for developers to implement complex cross-technological integrations between the data loading mechanisms and a program implemented in a BX DSL. This implementational simplification consequently reduces the cognitive load on developers.

We have used a data engineering problem to get a broader audience to extrapolate BX research in a practical direction. Data engineering is not the limit of BX potential, but rather a beginning. BX could hypothetically be used in program translation, software architecture, artificial intelligence - any area where a two-way mapping between two different data structures is observed. Therefore, we see \textit{Bifrons} as just the beginning of a far-reaching endeavor.

\bibliographystyle{IEEEtranN}
\bibliography{IEEEabrv,bibliography}
\pagebreak
\section*{Biography Section}
\vspace*{-\baselineskip}
\begin{IEEEbiography}[{\includegraphics[width=1in,height=1.25in,clip,keepaspectratio]{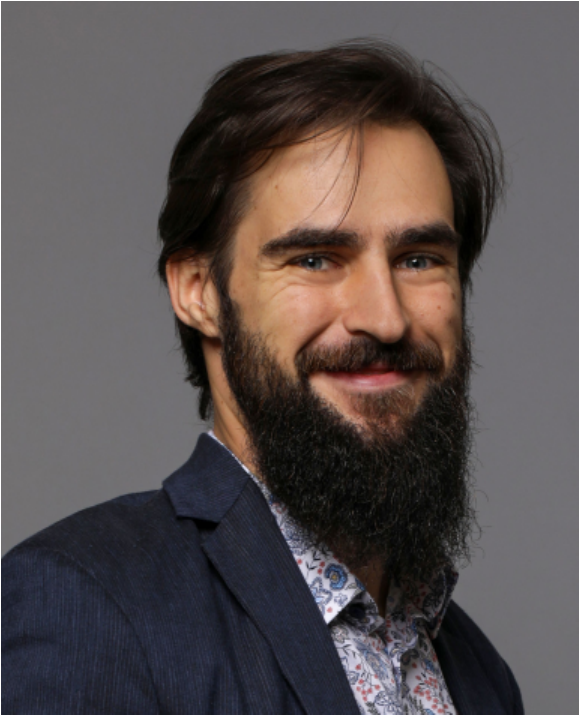}}]{Juraj Dončević}
received a Ph.D. degree in computer science from the Faculty of Electrical Engineering and Computing (FER), University of Zagreb. He is a teaching assistant at FER and does research related to software architectures, data integration, and bidirectionalization. He is a member of IEEE.
\end{IEEEbiography}
\vspace*{-\baselineskip}
\begin{IEEEbiography}[{\includegraphics[width=1in,height=1.25in,clip]{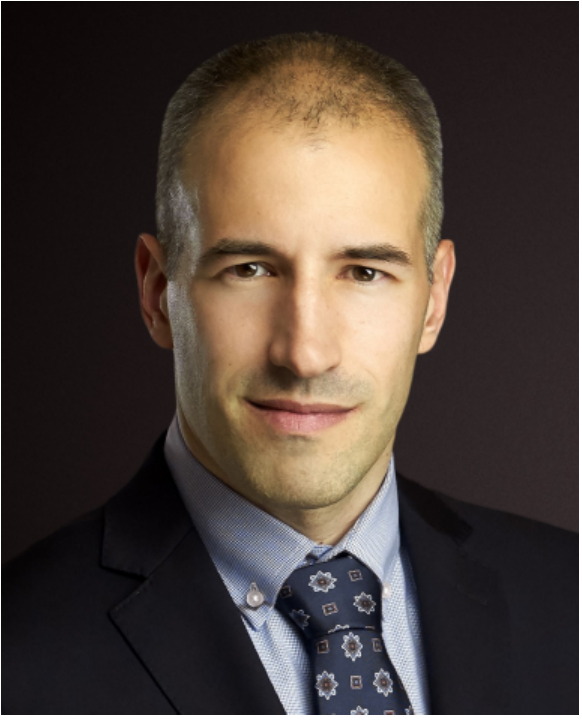}}]{Mario Brčić}
received a Ph.D. degree in computer science from the Faculty of Electrical Engineering and Computing, University of Zagreb, in 2015. He is currently an Associate Professor at the University of Zagreb. His research interests are at the crossroads of artificial intelligence and operations research. He is a member of IEEE, ACM, and AGI Society.
\end{IEEEbiography}
\vspace*{-\baselineskip}
\begin{IEEEbiography}[{\includegraphics[width=1in,height=1.25in,clip,keepaspectratio]{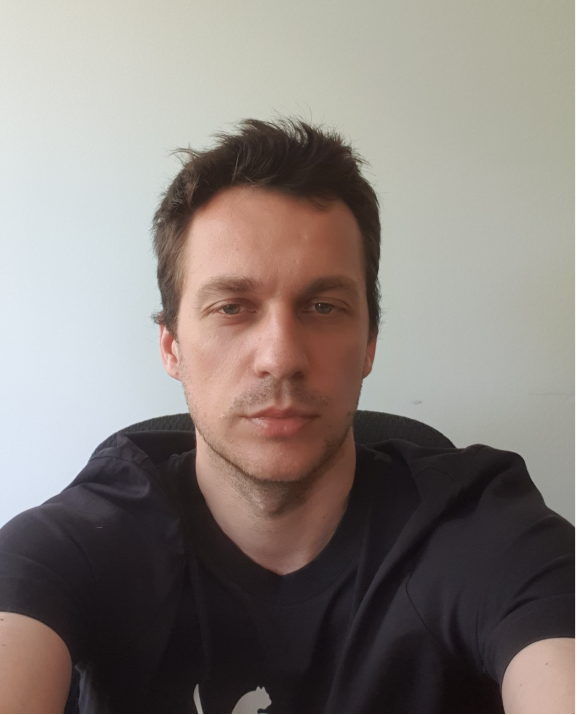}}]{Danijel Mlinarić}
received a Ph.D. degree in computer science from the Faculty of Electrical Engineering and Computing, University of Zagreb, in 2020. He is currently an Assistant Professor at the Department of Applied Computing. His research interests include software engineering, focusing on software evolution, program analysis, AI, and dynamic software updating. He is a member of IEEE.
\end{IEEEbiography}

\end{document}